\documentclass[a4paper,11pt]{article}

\pdfoutput=1
\usepackage{jheppub}

\usepackage{dsfont}
\usepackage{epsfig}
\usepackage{slashed}
\usepackage{bbold}
\usepackage{psfrag}
\usepackage{color}
\PassOptionsToPackage{caption=false}{subfig}
\usepackage{subfig}
\usepackage{multirow}
\usepackage{booktabs}
\usepackage{hyperref}
\usepackage{pstricks}

\usepackage{enumitem}
\usepackage[utf8]{inputenc}

\newcommand{\beqa}{\begin{eqnarray}}
\newcommand{\bea}{\begin{eqnarray}}
\newcommand{\eeqa}{\end{eqnarray}}
\newcommand{\eea}{\end{eqnarray}}
\newcommand{\beq}{\begin{equation}}
\newcommand{\eeq}{\end{equation}}
\newcommand{\bay}{\begin{array}}
\newcommand{\eay}{\end{array}}

\newcommand{\units}[1]{\mathrm{\; #1}}

\newcommand{\mysubsection}{\subsection}
\newcommand{\mysection}{\section}

\begin{document}
\preprint{YITP-SB-16-30}

\title{Searching for Dark Absorption with Direct Detection Experiments}

\author[a]{Itay M. Bloch,}
\emailAdd{itaybloc@mail.tau.ac.il}

\author[b]{Rouven Essig,}
\emailAdd{rouven.essig@stonybrook.edu}

\author[a,c]{Kohsaku Tobioka,}
\emailAdd{kohsakut@post.tau.ac.il}

\author[a]{Tomer Volansky,}
\emailAdd{tomerv@post.tau.ac.il}

\author[b]{Tien-Tien Yu.}
\emailAdd{chiu-tien.yu@stonybrook.edu}

\affiliation[a]{Raymond and Beverly Sackler School of Physics and Astronomy, Tel-Aviv University, Tel-Aviv 69978, Israel}
\affiliation[b]{C.N.~Yang Institute for Theoretical Physics, Stony Brook University, Stony Brook, NY 11794-3800}
\affiliation[c]{Department of Particle Physics and Astrophysics, Weizmann Institute of Science, Rehovot 7610001, Israel}

\abstract{
We consider the absorption by bound electrons of dark matter in the form of dark photons and axion-like particles, as well as of dark photons from the Sun, in current and next-generation direct detection experiments. Experiments sensitive to electron recoils can detect such particles with masses between a few eV to more than 10 keV. For dark photon dark matter, we update a previous bound based on XENON10 data and derive new bounds based on data from XENON100 and CDMSlite. We find these experiments to disfavor previously allowed parameter space. Moreover, we derive sensitivity projections for SuperCDMS at SNOLAB for silicon and germanium targets, as well as for various possible experiments with scintillating targets (cesium iodide, sodium iodide, and gallium arsenide). The projected sensitivity can probe large new regions of parameter space. For axion-like particles, the same current direction detection data improves on previously known direct-detection constraints but does not bound new parameter space beyond known stellar cooling bounds. However, projected sensitivities of the upcoming SuperCDMS SNOLAB using germanium can go beyond these and even probe parameter space consistent with possible hints from the white dwarf luminosity function.  We find similar results for dark photons from the sun. For all cases, direct-detection experiments can have unprecedented sensitivity to dark-sector particles.
}

\maketitle
\flushbottom

\mysection{Introduction}

Significant experimental evidence and theoretical considerations suggest that  the Standard Model (SM) of particle physics is incomplete.  
In particular, there is compelling evidence for the existence of dark matter (DM).  
Determining the identity of the DM particle is one of the most important problems in particle physics today.   

Many DM candidates exist, with Weakly Interacting Massive Particles (WIMPs) being the most studied. 
Numerous direct-detection experiments probe for the elastic scattering of WIMPs (and other DM candidates) off ordinary matter~\cite{Cushman:2013zza}. 
However, it is entirely plausible that DM is composed of a (pseudo)scalar or a vector boson that can be {\it absorbed} 
in a material, thereby depositing its full kinetic and rest-mass energy.  
DM in the form of axions, axion-like particles (ALPs), and dark photons can have this property, see e.g.~\cite{Pospelov:2008jk}.  
This paper focuses on the search of such DM candidates with existing and upcoming 
direct-detection experiments. 
For other types of searches see 
e.g.~\cite{Jaeckel:2007ch,Ahlers:2007rd,Wagner:2010mi,Povey:2010hs,Horns:2012jf,Parker:2013fba,Betz:2013dza,Dobrich:2014kda,Chaudhuri:2014dla,Graham:2014sha,Kouvaris:2016afs}. 

The axion was originally introduced as a solution to the strong CP problem~\cite{Weinberg:1977ma,Wilczek:1977pj}.   
This pseudoscalar may couple in a model-dependent manner to the SM axial current,  
$\partial^\mu a J_\mu^A / f_a$, where $f_a$ is the axion decay constant and indicates the scale at which the 
Peccei-Quinn~\cite{Peccei:1977hh} symmetry is broken.  
More generally, the spontaneous breaking of any global symmetry combined with a small explicit breaking can lead to 
low-mass ALPs with a similar interaction. 
This interaction allows the axion or ALP to be absorbed in a target material, rather than scattering off it. 
The case of absorption in bound electrons, known as the axioelectric effect, has been originally suggested in~\cite{Dimopoulos:1985tm,Avignone:1986vm}.  Numerous direct-detection experiments have conducted searches assuming that the ALP constitutes the 
galactic DM or is produced in the Sun, including 
 CoGeNT~\cite{Aalseth:2008rx}, CDMS~\cite{Ahmed:2009ht}, EDELWEISS~\cite{Armengaud:2013rta}, XENON100~\cite{Aprile:2014eoa}, and  KIMS~\cite{Yoon:2016ogs}.   
 Here we show that other existing data can improve on these bounds and provide projections for future experiments.  
We find that future experiments could go beyond the strong stellar constraints. 

Another simple DM candidate, present in many extensions of the SM, is the dark photon, which we denote as $A'$. 
The $A'$ is a massive vector boson that can couple weakly to ordinary matter through kinetic 
mixing~\cite{Holdom:1985ag,Galison:1983pa,ArkaniHamed:2008qn,Pospelov:2007mp,Pospelov:2008jd}. 
Significant ongoing experimental effort searches for an $A'$, see e.g.~\cite{Jaeckel:2010ni,Hewett:2012ns,Essig:2013lka}. 
An intriguing possibility, however, is that the $A'$ itself is sufficiently stable to play the role of the DM particle~\cite{Nelson:2011sf,Arias:2012az,Graham:2015rva}.   
In this case, relic $A'$s may be absorbed by ordinary matter in direct-detection experiments, just like an ALP.  
Initial studies of this possibility were presented in~\cite{An:2014twa}.  
We derive new direct-detection bounds that are more constraining than existing limits, and provide projected sensitivities. 

We focus on the sensitivity gain achieved when lowering the threshold of direct-detection experiments.  
Traditionally, these experiments have been optimized to probe WIMPs with masses above $\mathcal{O}$(GeV), 
requiring sensitivity to nuclear recoil energies above $\mathcal{O}$(keV).  
However, many viable DM candidates have masses well below the GeV-scale, which can only be probed with new 
techniques~\cite{Essig:2011nj}.  
DM scattering off electrons is one such technique suggested in~\cite{Essig:2011nj}.  
Not only does this probe sub-GeV DM scattering, but a lower ionization threshold translates directly into a lower mass threshold 
for absorbing (nonrelativistic) DM. 

XENON10's sensitivity to single electrons~\cite{Angle:2011th} allows them to probe the lowest electron recoil energies,
but without distinguishing signal from background events.  
This data was used to place a constraint on DM as light as a few MeV scattering off electrons~\cite{Essig:2012yx} and on $A'$ DM as light as the xenon ionization energy of 12.1~eV~\cite{An:2014twa}.  
We revisit the latter limit below.   
More recently, the XENON100 experiment published a search for events with four or more electrons~\cite{Aprile:2016wwo}.  
We show that this data improves constraints on sub-keV ALPs and $A'$ DM.  
We also show that a CDMSlite dataset sensitive to ${\cal O}(20)$ electrons~\cite{Agnese:2015nto} sets 
the strongest constraints for some $A'$ DM masses.  

Future experiments are expected to achieve improved sensitivity to even lower DM masses by lowering the threshold down to 
$\mathcal{O}$(eV) and having a reduced background.  
A key to this improvement is the use of semiconductors or scintillators, which have band gaps of $\mathcal{O}$(few eV).
The sensitivity gained by developing ultra-sensitive detectors able 
to probe one or a few electrons in semiconductors (e.g.~SuperCDMS~\cite{Agnese:2015nto}, 
DAMIC~\cite{Moroni:2011xs,Aguilar-Arevalo:2016ndq}), scintillators, or two-dimensional targets has been studied in the context of 
sub-GeV DM scattering in~\cite{Essig:2011nj,Essig:2015cda,Graham:2012su,Lee:2015qva,Hochberg:2016ntt,Derenzo:2016fse}.
Below we study the prospects of such experiments to detect or constrain ALPs and $A'$s. 
Superconducting targets may probe even lower masses, either by DM scattering~\cite{Hochberg:2015fth,Hochberg:2015pha} or 
DM absorption~\cite{Hochberg:2016ajh}, in the future.  
Moreover, while beyond the scope of this paper, chemical-bond breaking could be used to probe nuclear recoils down to 
10~MeV~\cite{Essig:2016crl,cc}, while superfluid helium could probe nuclear recoils down to keV~\cite{Schutz:2016tid}.  
These would probe different couplings from the models considered here.  

Finally, axions, ALPs, and dark photons may exist without constituting a significant fraction of the DM energy density.  
If sufficiently light, they would be produced in the Sun and can again be absorbed in direct-detection experiments. 
We will discuss constraints from existing experiments and provide projections for upcoming experiments.  
Since ALPs are dominantly produced in the Sun with $\sim$keV energies, no gain is achieved by experiments lowering their threshold.
Instead, we focus on dark photons, which are produced also at low energies, so that lower thresholds could potentially probe beyond existing constraints. 

\mysection{Axion-Like Particles}
The original ``QCD'' axion has a strict relation between its mass, $m_a$, and decay constant, $f_a$.   More generally, however,  ALPs 
may be described by independent $m_a$ and $f_a$. 
Moreover, ALPs have model-dependent coupling to gluons, photons, and SM fermions.  
We are interested here in possible interactions with electrons, defined minimally by 
\begin{eqnarray}
\label{eq:1}
{\cal L}_a = \frac{1}{2}\partial_\mu a\partial^\mu a - \frac{1}{2} m_a^2 a^2 + i g_{aee} a \bar e \gamma_5 e\,.
\end{eqnarray}
In the absence of additional couplings, the model dependence in $g_{aee}$ can be absorbed in $f_a$ and the two are related by
$g_{aee} = 2 m_e/f_a$.
We show existing bounds and projections in the $g_{aee}$ versus $m_a$ parameter space.  

The ALP interaction with electrons is responsible for the ``axioelectric'' effect~\cite{Dimopoulos:1985tm,Avignone:1986vm,Pospelov:2008jk,Derevianko:2010kz}.  In analogy to the photoelectric effect, axions may be absorbed by bound electrons, ionizing them.  
Direct detection experiments search, in part, for ionized electrons and therefore offer an opportunity to search for ALPs via 
their coupling to electrons.     

The ALP absorption cross section, $\sigma_{\rm AE}$, may be related to the photoelectric 
absorption cross section, $\sigma_{\rm PE}$, as 
\begin{eqnarray}
\label{eq:2}
\sigma_{\rm AE}(E) v_a \simeq \sigma_{\rm PE}(E)  \frac{3}{4}\frac{g_{aee}^2}{4\pi \alpha_{\rm EM}} \frac{E^2}{m_e^2}\left(1-\frac{1}{3}v_a^{2/3}\right)\,,
\end{eqnarray}
where $E$ and $v_a$ are the ALP's energy  and velocity, respectively~\cite{Dimopoulos:1985tm,Avignone:1986vm,Pospelov:2008jk}.  
We infer that the rate of absorption of non-relativistic ALP DM on an electron ($n_a \sigma_{\rm AE} v_a$, 
where $n_a$ is the ALP number density) has negligible dependence on $v_a$.  
The absorption spectrum is thus very narrow, helping to differentiate signal from background.

We take $\sigma_{\rm PE}(E)$ from~Henke et al.~\cite{Henke:1993eda, HenkeDatabase} and the \emph{Handbook of Optical Constants of Solids}~\cite{PALIK1985429,POTTER1985465,EDWARDS1985547,ELDRIDGE1985853}.  
Discrepancies of $\mathcal{O}(50\%)$ exist between different theoretical predictions for $\sigma_{\rm PE}(E)$ 
below $\sim\! 300$~eV. 
The main sources of theory uncertainties arise from approximating the correlation between different electrons. 
We use the Henke calculations, which are favored by experimental data~\cite{HenkeExp}, for energies above 30~eV. Below 30 eV, we use the measurements compiled in ~\cite{PALIK1985429,POTTER1985465,EDWARDS1985547,ELDRIDGE1985853}.


\mysubsection{ALPs as Dark Matter}

Axions and ALPs may play the role of DM.  The QCD axion is produced either thermally or through the misalignment mechanism and the decay of cosmological defects (for a review see e.g.~\cite{Sikivie:2006ni,Marsh:2015xka}).   
By tuning the initial misalignment angle or through late-time entropy dumping, axion DM has masses $\lesssim \!20 \, \mu\mathrm{eV}$.   
Non-thermal production through defects may allow for heavier axions; quite generally though, its mass must be $\lesssim\,$eV 
for it to be cold DM. 
Hot axions also require masses $\lesssim\,$eV. 

ALPs could obtain the correct relic density in a wider mass range, via the misalignment mechanism, or by thermal or non-thermal production. 
It is thus interesting to constrain ALPs without assuming a specific production mechanism.   
The ALP absorption rate on electrons is 
\begin{eqnarray}
\label{eq:3}
R_{\rm ALPs} = 1.9\times 10^{19} \units{kg^{-1} day^{-1}} \frac{g_{aee}^2}{A}  \left(\frac{m_a}{\units{keV}}\right)\left(\frac{\sigma_{\rm PE}}{\units{bn}}\right) \,,
\end{eqnarray}
for a local DM energy density of $\rho_{\rm DM} = 0.4\units{GeV/cm^3}$.

\mysubsection{ALPs from the Sun}

Independent of their relic density, ALPs may be produced in the Sun.   
The flux has been calculated in~\cite{Redondo:2013wwa} and is largest at $\mathcal{O}$(keV) 
energies. 

Solar ALPs can be detected in direct-detection experiments~\cite{Dimopoulos:1985tm,Avignone:1986vm}.
Limits on $g_{\rm aee}$ versus the ALP mass have been published by several direct detection experiments, among the strongest of which is 
the XENON100 collaboration analysis that used an S1 (primary scintillation) trigger with a 2 keV threshold~\cite{Aprile:2014eoa}.  
The more recent S2-only analysis~\cite{Aprile:2016wwo}, which searched for events with an ionization-only (S2) signal, has a 
much lower experimental threshold and is a larger dataset than that used in~\cite{Aprile:2014eoa}. However, the solar ALP flux drops rapidly at low energies, while the S2-only analysis includes more background events due to the lower threshold. Therefore, the S2-only dataset produces a weaker bound than found in~\cite{Aprile:2014eoa}.
Future xenon-based experiments can improve on these limits due to their large exposures and low expected 
backgrounds (see e.g.~\cite{Avignone:2009ay},  although exposures $\gtrsim\!100$~ton-years are required to probe beyond 
white-dwarf cooling limits).  
Conversely, lowering the experimental threshold without a significant gain in exposure over XENON100~\cite{Aprile:2014eoa}
only provides a negligible increase in sensitivity due to the  rapidly falling solar flux below $\mathcal{O}$(keV) energies.  
Since this paper is focused on discussing the potential gain from low-threshold experiments, we do not present the solar ALP limits.

\mysection{Dark photons}

The $A'$ is a hypothetical massive vector boson of a broken (dark) gauge group $U(1)_D$ 
that may kinetically mix with the SM hypercharge~\cite{Holdom:1985ag,Galison:1983pa}. 
At low energies, this mixing is dominantly between the $A'$ and the SM photon. 
The relevant interactions are 
\bea
\mathcal{L} \supset - \frac{1}{4}\,F'^{\mu\nu} F'_{\mu\nu} - \frac{\epsilon}{2}\,F^{\mu\nu} F'_{\mu\nu} + \frac{1}{2}\,m_{A'}^2 A'^\mu A'_\mu \,,
\label{eq:aprime_lagrangian}
\eea
where $\epsilon$ is the kinetic-mixing parameter, $m_{A'}$ is the $A'$ mass, and $F^{\mu\nu}$ is the SM photon field 
strength.  
The mass $m_{A'}$ can have two origins: a St\"uckelberg mass, which predicts  no additional degrees of freedom, 
or a Higgs mechanism, where the $U(1)_D$ is broken spontaneously by an extra Higgs field. 
This distinction is irrelevant for $A'$ DM detection, but important for detecting $A'$ from the Sun.  
For the latter, we only consider a St\"uckelberg mass.   

\mysubsection{Dark photons as Dark Matter}

For sufficiently small $\epsilon$ and $m_{A'}$ below twice the electron mass, the $A'$ decay lifetime can be longer than the 
age of the Universe, allowing for the $A'$ to constitute all the DM.  
Various production mechanisms exist, see e.g.~\cite{Nelson:2011sf,Arias:2012az,Graham:2015rva}.   
The absorption of $A'$ DM can be modeled as the absorption of a massive non-relativistic particle with coupling $e\epsilon$ 
to electrons. 
We can then write the $A'$ absorption cross-section in terms of $\sigma_{\rm PE}$,
\beq
\sigma_{A'}(E_{A'}=m_{A'})\,v_{A'}\simeq \epsilon^2 \sigma_{\rm PE}(E=m_{A'}) \,,
\eeq
where $v_{A'}$ is the dark photon velocity \cite{Pospelov:2008jk}. The resulting absorption rate is given by
(for $\rho_{\rm DM} = 0.4$~GeV/cm$^3$) 
\beq
{\rm{Rate~per~atom}} \simeq \frac{\rho_{\rm DM}}{m_{A'}}\times \epsilon^2\sigma_{\rm PE}(E=m_{A'}) \,.
\eeq

We include in-medium effects.  These are especially important for the detection of solar $A'$s, 
so we discuss them in the next subsection.   
For $A'$ DM, the absorption rates are affected only by $\lesssim 10\%$ for $m_{A'} \gtrsim100$~eV, 
although for $m_{A'}\lesssim 100$~eV, the rate for xenon (germanium, silicon) is changed by a factor of $\sim\,$0.2--2.0 (1.0--1.15, 1.0--1.8).

\mysubsection{Dark photons from the Sun}

Direct detection experiments are sensitive to $A'$s from the Sun~\cite{An:2013yua,Redondo:2008aa}. 
The detection of these solar $A'$s is analogous to that of $A'$ DM, but here the in-medium effects of the $A'$ 
in the Sun may significantly affect production and detection. 
We focus on the St\"uckelberg case, and summarize the effect as described in~\cite{An:2013yfc,An:2013yua}. 

The matrix element for $A'$ absorption in a medium is 
\beq
\mathcal{M}(A'_{T,L}+i\to f)=-\frac{\epsilon m_{A'}^2}{m_{A'}^2-\Pi_{T,L}}\left[e J_{\rm{em}}^\mu\right]_{fi}\epsilon_\mu^{T,L}\,.
\eeq
Here $i$ and $f$ denote the initial and final electron state and $\Pi_{T,L}$ are the transverse (T) and longitudinal (L) in-medium polarization functions defined by 
\beq
e^2\langle J_{\rm{em}}^\mu,J_{\rm{em}}^\nu\rangle=\Pi_T\epsilon_i^{T\mu}\epsilon_i^{T\nu}+\Pi_L\epsilon^{L\mu}\epsilon_i^{L\nu}\,, 
\eeq
which encode the correlation function inside the medium.
The absorption rate of the transverse and longitudinal modes of a solar $A'$ in the detector frame is 
\begin{align}
\Gamma_{T,L}^{\rm detect}=\frac{\epsilon^2_{T,L} \rm{Im}~\Pi_{T,L}}{\omega}\,,
\end{align}
where the effective mixing angle is 
\beq
\epsilon_{T,L}^2=\frac{\epsilon^2 m_{A'}^4}{(m_{A'}^2-\rm{Re}~\Pi_{T,L})^2+(\rm{Im}~\Pi_{T,L})^2}\,.
\label{eq:effective_epsilon}
\eeq	
For an isotropic and non-magnetic material with a refractive index $n_{\rm refr}$, one has 
\begin{align}
\Pi_T=\omega^2 (1-n_{\rm refr}^2), \quad \Pi_L=(\omega^2-|\vec q|^2) (1-n_{\rm refr}^2)\,,
\end{align}
where $q=(\omega,\vec q)$ is the $A'$ 4-momentum, with 
\begin{align}
n_{\rm refr}=1-\frac{r_0}{2\pi}\lambda^2 \sum_A n_A(f_1^A+if_2^A)\,.
\end{align}
Here $n_A$ denotes the density of atoms of type $A$, $\lambda=2\pi\hbar c/\omega$ is the wavelength of a photon with energy $\omega$, and $r_0$ is the classical electron radius, $e^2/(m_ec^2)=2.82\times 10^{-15}$~m. 
The atomic scattering factors $f_{1,2}$ are given in~\cite{Henke:1993eda, HenkeDatabase} 
The database~\cite{HenkeDatabase} has $f_1$ above 30~eV and $f_2$ above 10~eV. 
To obtain $f_1^{\rm Xe}$ for energies below 30 eV, we use ${\rm Re}[n_{\rm refr}]$ in Fig.~3 of~\cite{An:2014twa}. 
 
The flux of solar dark photons is given by 
\begin{align}
\frac{d\Phi_{T,L}}{d\omega}=\frac{1}{4\pi d_\odot^2}\int_0^{R_\odot}\! dr 4\pi r^2  \ \frac{d\Gamma_{T,L}^{prod,V}}{d\omega dV}\,, \label{RatetoFlux}
\end{align}
where $d_\odot=1$~AU is the Earth-Sun distance and $R_\odot=6.96\times 10^8$~m is Sun's radius.
We consider the resonant production of the transverse and longitudinal modes, which gives the dominant  component of the flux.
The spectrum of the total number of events is
\bea
\frac{dN_{events}}{d\omega}&=&VT \frac{\omega}{\sqrt{\omega^2-m_{A'}^2}}   {\rm Br} 
 \left(\frac{d\Phi_{T}}{d\omega}\Gamma_{T}^{detect}+\frac{d\Phi_{L}}{d\omega}\Gamma_{L}^{detect}\right),
\eea
where $V,T$ are the detector's fiducial volume and exposure time, respectively. 
The ratio of the ionization rate to the absorption rate, Br, is unity for the energies of interest. 
The spectrum has a sharp rise at the mass threshold because of the resonant production of the longitudinal mode.

\mysection{Analysis of Current and Future Experiments}

In this section, we describe how we derive bounds on ALPs and $A'$s using existing data from XENON10~\cite{Angle:2011th}, 
XENON100~\cite{Aprile:2016wwo}, and CDMSlite~\cite{Agnese:2015nto}, 
and how we derive projections for possible future experiments.  

The low-mass reach of an experiment for ALP DM or $A'$ DM absorption depends on the type of material used as a target.  
Xenon-based targets have a low-mass reach of 12.1~eV, while semiconductor- or scintillator-based experiments 
have a band gap, which is lower by about an order of magnitude, 
allowing such experiments to lower sensitivity down to $\sim$eV DM masses. 

\mysubsection{Signal Region}

\noindent 
To extract optimal limits, we need to take into account the spectral shape of the signal in the detector, which depends not only on the 
detector energy resolution.  
In particular, an electron that absorbs an ALP or $A'$ (the `primary' electron) is ejected from the atom, and its resulting 
kinetic energy (total minus binding) is converted to ionizing additional atoms, which can lead to several additional (`secondary') electrons.  
The number of secondary electrons produced fluctuates from event to event.  
Their variance and average number can be predicted in a given experiment, either by modeling 
the secondary interactions or through measurements.  
For a highly energetic primary electron, this variance usually dominates over the detector resolution.  

We estimate below, either directly from the experimental data or from theoretical expectations, 
the variance and detector resolution (adding them in quadrature if needed), and smear the theoretical spectrum accordingly.   
To derive a conservative limit, we choose a single bin containing 95\% of the signal and demand that it is less than 
the data at 90\% C.L.. 
For the solar $A'$ signal, we use a slightly simpler procedure, since the signal is distributed over a wider energy range: 
we first calculate the energy range (from $m_{A'}$ upwards) that contains 95\% of the signal, then smear the first and last bins, and 
require the signal to be less than the data at 90\% C.L.~in the resulting (smeared) energy range. 
The energy range we can use is limited by the ionization threshold in xenon ($>\!12.1$~eV) and by 
the band gap for the other materials (see below Eq.~(\ref{eq:2})). Note that the atomic data for NaI does not extend below below 30 eV, and so we do not go below 30 eV. 

\mysubsection{Current Experiments}

\begin{figure}[t]
	\centering
	\includegraphics[width=0.45\textwidth]{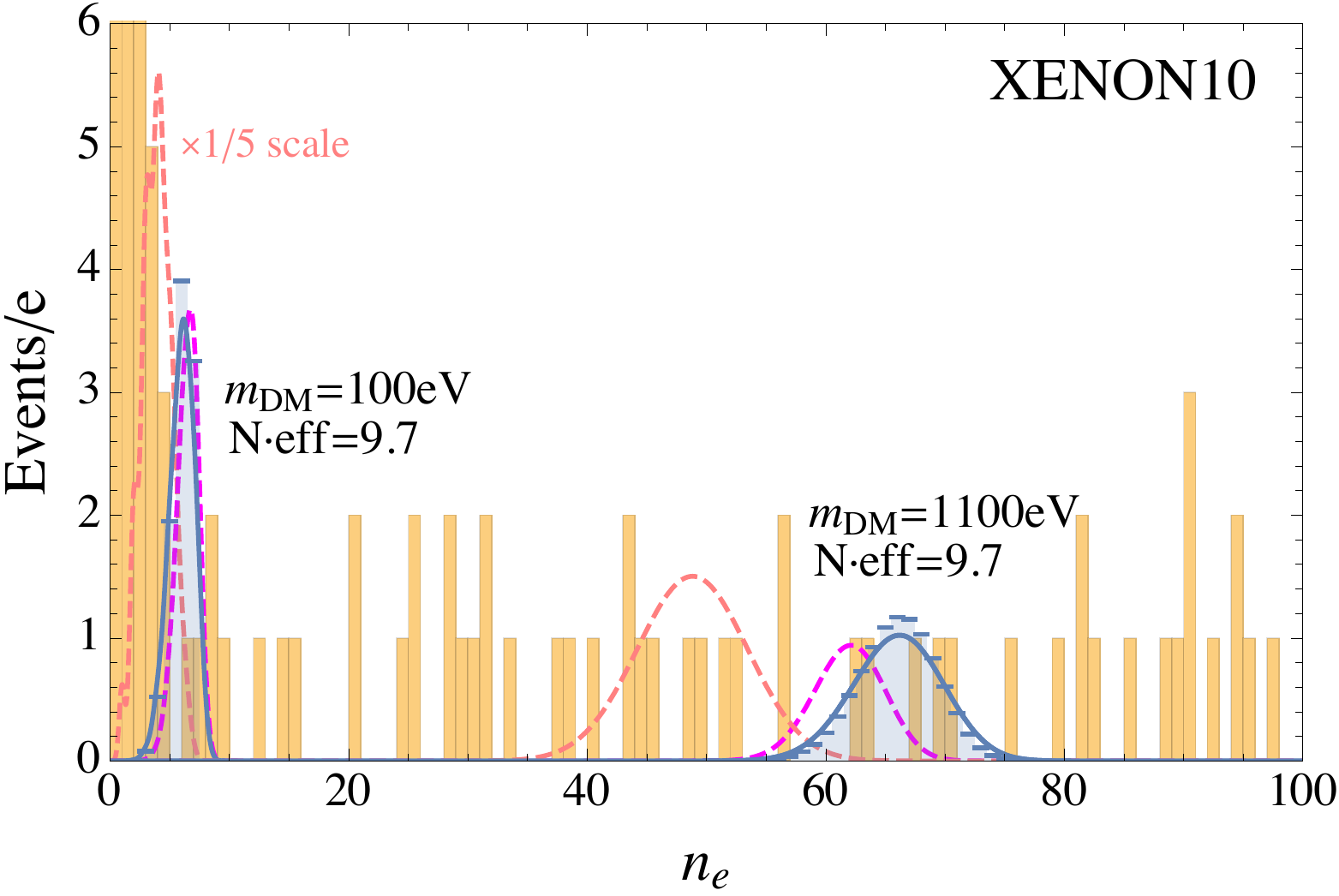}
	\includegraphics[width=0.47\textwidth]{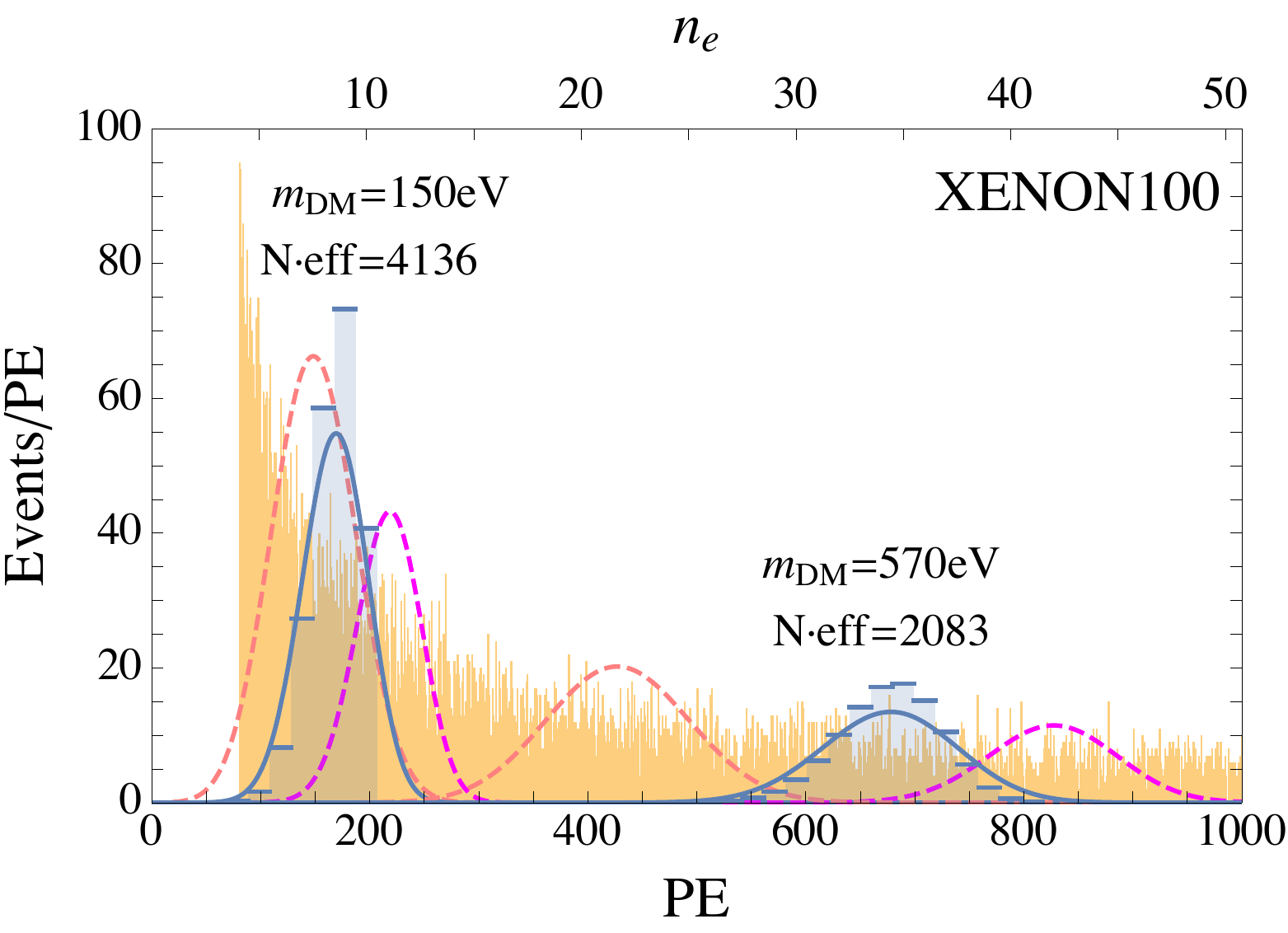}
	\caption{Ionization (S2-only) data from XENON10~\cite{Angle:2011th} ({\bf left}) and XENON100~\cite{Aprile:2016wwo} ({\bf right}). 
	Top plot shows the data as the number of observed electrons, while bottom plot shows the data as the number of 
	photoelectrons (PE) on the bottom axis and the corresponding number of electrons (1e$^-$=19.7PE) on the top axis.  
	Blue gaussian lines on each plot show two examples of the expected observed signal shape for two different 
	DM (ALP or $A'$) masses.  The blue histogram shows the signal distribution in terms of the number of electrons produced in the 
	detector.  
	Varying the secondary ionization model produces different signal shapes and thus different limits.  
	We show the signal shape in pink and magenta that produces, respectively, the worst and best limit for the same two DM 
	masses when varying the secondary ionization model. Numbers next to the blue curves give the 90\% C.L.~upper bound 
	times efficiency on the number of signal events for those masses.  
	}
	\label{fig:XENONspectra}
\end{figure}

\noindent We consider the following existing datasets:
\begin{itemize}[leftmargin=*]
\item {\bf XENON10:}
The XENON10 experiment uses xenon as the target material and presented 
results using 15 kg-days of data with an S2 (ionization signal) trigger threshold set to a single electron~\cite{Angle:2011th}.   
Previous work used these data to derive conservative bounds on sub-GeV DM~\cite{Essig:2012yx},  
solar $A'$~\cite{An:2013yua}, and $A'$ DM~\cite{An:2014twa}.  
In the latter two, the signal was required to be less than the {\it total} number of observed S2 events 
with $\le \!80$ electrons, without taking into account spectral information. 
Here we use the spectral information to place more stringent bounds, except 
near threshold ($\sim\!15-30$~eV), 
 where the bound is in reality slightly weaker.  
We take the signal efficiency times acceptance to be 0.92 for events with $\ge 2$ electrons.  
For single electrons, the efficiency is only about 0.5, and we require the signal event rate to be 
less than 23.4 events/kg/day~\cite{Essig:2012yx}.  

The number of secondary ionization electrons is challenging to calculate precisely, but we model it as discussed in~\cite{Essig:2012yx}.  
Each electron produces $\sim 27.0\pm 6.2$ photoelectrons (PE), which are the actual signal seen by the XENON10 detector and 
defines the detector resolution.  
In~\cite{Angle:2011th}, the data has already been converted from the number of PEs to the number of electrons.  
In Fig.~\ref{fig:XENONspectra} (top), we reproduce the spectrum of events (orange histogram) shown in Fig.~2 of~\cite{Angle:2011th}. 
We also show (in blue) a hypothetical DM ALP/$A'$ signal, with the normalization chosen at the 90\% C.L.~limit, at two different masses 
(100~eV and 1100~eV). 
The pink and magenta gaussian lines show the signal shape that produces the worst and best  
limit, respectively, for the same two DM masses  when varying the secondary ionization model.  
In Fig.~\ref{fig:solarA'spectrum}, we show examples of solar $A'$ spectra.  
Since the signal spans a wider energy range, the uncertainty in the secondary ionization modeling should be small, and we do not include the uncertainty.  
\item {\bf XENON100:} The XENON100 collaboration presented S2-only data with events down to 80 PE, 
corresponding to about 4~electrons, in~\cite{Aprile:2016wwo}.  
The data listing the number of PE for each event is reproduced 
in Fig.~\ref{fig:XENONspectra} (bottom), where we bin the data in 1~PE bins. 
The mean number of PE generated by one electron is 
$19.7\pm0.3$, with a width of $6.9 \pm 0.3$ PE/electron~\cite{Aprile:2016wwo}. 
Our signal calculation includes the acceptance and trigger efficiencies from~\cite{Aprile:2016wwo}.  
As for XENON10 above, we show a hypothetical DM ALP/$A'$ signal in blue at 
two different masses (150~eV and 570~eV).  
The pink and magenta gaussian lines again show, for the same two DM masses, 
the signal shape that produces the worst and best  limit when varying the secondary ionization model. 
We derive the bound on solar $A'$s as discussed above. 
The bound on sub-GeV DM scattering off electrons based on this data will be presented in~\cite{Essig:2017kqs}.  

\begin{figure}[t]
	\centering
	\includegraphics[width=0.57\textwidth]{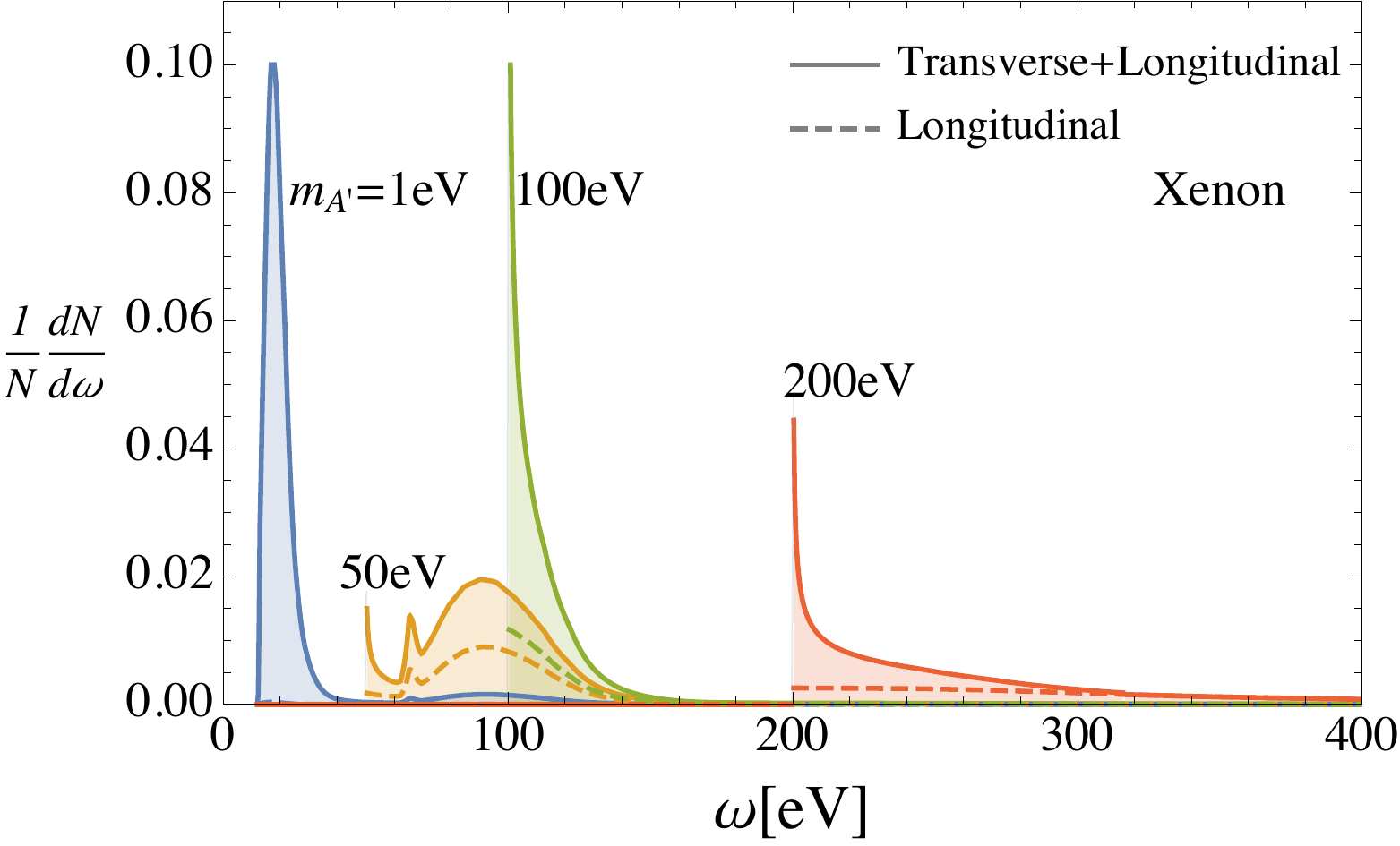} 
	\caption{The normalized spectra of the total number of events for dark photons ($A'$) from the Sun in a xenon detector.  The combined transverse and longitudinal modes ({\bf solid}) and the longitudinal mode alone 
	({\bf dashed}) are shown for $m_{A'}=1, 50, 100,~\rm{and}~200$ eV. 
	The spectrum of the $m_{A'}=1$ eV case is cut off by the xenon ionization energy, $12.1$ eV.
	\vspace{-4mm}
	}
	\label{fig:solarA'spectrum}
\end{figure}

\item {\bf CDMSlite:} The SuperCDMS collaboration reported results with a threshold of 56~eV from a low-threshold ``CDMSlite'' run 
in~\cite{Agnese:2015nto}.  
To derive a bound, we take the data  from Fig.~3 of~\cite{Agnese:2015nto} and apply all efficiencies and cuts 
 (note that Fig.~3 is not corrected for the trigger efficiency). 
We treat the energy range 56 eV--140 eV as a single bin, taking the data from Table~I. 
The total resolution of the signal (combining the experimental resolution and fluctuations in the number of secondary electrons) 
can be inferred from the observed electron capture peaks. 
Thus we adopt $\sigma$=18.24~eV (30.68~eV, 101~eV) for the mass ranges [56~eV, 160~eV] ([160~eV, 1.3~keV],  [1.3~keV, 10.37~keV]). 
\end{itemize}

\mysubsection{Future Experiments}

\noindent 
Future experiments are expected to have lower thresholds, allowing them to probe lower DM masses.  
To derive projections, we take into account expected physics backgrounds, but assume zero detector dark counts, 
i.e.~no spontaneous creation of electron-hole pairs that mimic a signal. 
\begin{itemize}[leftmargin=*]
\item {\bf SuperCDMS SNOLAB HV Ge and Si:} 
The next-generation SuperCDMS experiment will be deployed at SNOLAB.  
A high-voltage (HV) version will attempt to use a large bias voltage to achieve the same Luke-Neganov-phonons amplification 
of an ionization signal as the previous CDMSlite versions.  
Ultra-sensitive phonon detectors may achieve sensitivity down to single electrons, allowing the threshold to be given by the 
band gap (0.67~eV for Ge, 1.1~eV for Si).  
The projected exposures are 20 (10)~kg-years for germanium (silicon)~\cite{Cushman:2013zza}.  
Our projections assume the following backgrounds: 
\begin{itemize}[leftmargin=*]
\item {\bf Germanium:}  We use a preliminary estimate by the SuperCDMS collaboration 
in~\cite{GolwalaTalk}.  
This estimate includes backgrounds from tritium $\beta$-decay (a flat background on a log-log plot 
between $\sim 1$~eV to a few keV), various x-ray lines produced by cosmogenic activation, and solar neutrinos 
(for exposures of $\lesssim$ few kg-years, these are only relevant at electron recoil energies $\lesssim 10$~eV).  

\begin{table}[t]
\begin{center}
\begin{tabular}{|c||c|c|c|}
\hline
Element & $\varepsilon_e$ [eV] & $E_g$ [eV] & $\sigma_{ee}$ at $E_e = 1$~keV [eV] \\ \hline
Ge & 2.9 & 0.67 & 19.4 \\
Si & 3.6 & 1.1 & 21.7 \\
NaI & 17.7 & 5.9 & 47.8 \\
CsI & 19.2 & 6.4 & 50.0 \\
GaAs & 4.2 & 1.52 & 23.4 \\
\hline
\end{tabular}
\caption{Mean energy to create an electron-hole pair ($\varepsilon_e$), band gap energy ($E_g$), and signal width ($\sigma_{ee}$) 
at ionization energies of 1~keV from Eq.~(\ref{eq:sigmaee}) in various semiconductors (Si, Ge) 
and scintillators (NaI, CsI, GaAs). 
\label{tab:Eg}
	\vspace{-4mm}
}
\end{center}
\end{table}%

We assume that an electron in germanium absorbs a DM particle of mass $m_{\rm DM}$, which must be larger than the 
band gap $E_g$.  
The kinetic energy ($E_e -E_g = m_{\rm DM}-E_g$) of this primary electron is converted into additional electron-hole pairs.  
The mean total number of electron-hole pairs created is  
\begin{equation}
\langle Q(E_e)\rangle = 1 + \lfloor (E_e - E_g)/\varepsilon_e \rfloor \, ,
\label{eq:linear-ionization-response}
\end{equation}
where $\varepsilon_e$ is the mean energy per electron-hole pair and $\lfloor y \rfloor$ rounds $y$ down to the nearest integer. 
Fluctuations around the mean are given by the Fano factor, $F \equiv \sigma_Q^2/\langle Q \rangle$, where 
$\sigma_Q^2$ is the variance.
To calculate the width in the signal's energy, we add in quadrature the energy resolution of the phonon detectors, $\sigma_t$, 
and the fluctuations in the number of secondary electron-hole pairs, 
\bea\label{eq:sigmaee}
\sigma_{\rm ee} = \varepsilon_e~\sqrt{\frac{\sigma_t}{q_e V_b} + F\, \langle Q(E_e)\rangle} \,.
\eea
Here, $V_b$ is the bias voltage, so one electron-hole pair will produce a total phonon energy of $q_e V_b$, where $q_e=1$ is the 
electron's charge.  
For DM absorption, where $E_e = m_{\rm DM}$, this equation becomes 
\bea
\frac{\sigma_{\rm ee}}{m_{\rm DM}} \simeq \frac{\sqrt{\frac{\sigma_t}{q_e V_b} + F \langle Q(E_e)\rangle}}{\langle Q(E_e)\rangle -1 + \frac{E_g}{\varepsilon_e}} \,.
\eea
For large enough $Q$, the signal shape is dominated by the Fano factor. 

For germanium (and all other elements below), we take $F=0.13$~\cite{Lowe1997, Lepy2000}. While $F$ varies less than 50\% over almost two orders in magnitude in energy~\cite{Lowe1997}, measurements of $F$ at energies below $\sim 200$~eV 
would be desirable.   
We take $\sigma_t=10$~eV, although the precise value makes a difference only at the lowest masses above the 
band gap when only $\mathcal{O}({\rm few})$ electrons are created. 
We set $V_b = 100~V$, and use the values of $\varepsilon_e$ and $E_g$ for germanium (and other elements) 
in Table~\ref{tab:Eg}~\cite{ExptGaps,Klein:1968}.
We list $\sigma_{ee} (1~{\rm keV})$ as an example of the signal widths for various elements.   
\item {\bf Silicon:}  Silicon does not have the same low-energy x-ray lines produced by cosmogenic activation. 
However, $^{32}$Si is a relatively large contaminant and a low-energy $\beta$-emitter.  Together with tritium, this produces a flat 
background at energies $\lesssim\! 10$~keV, which we take to be 350~events/keVee/kg/year based on preliminary 
estimates by the SuperCDMS collaboration~\cite{RobinsonTalk}. 
We take $\sigma_t=10$~eV, although the precise value is again only important just above 
the band gap.  See Table~\ref{tab:Eg} for the values of $\varepsilon_e$, $E_g$, and $\sigma_{ee} (1~{\rm keV})$ 
for silicon.\footnote{Future versions of the DAMIC experiments~\cite{Barreto:2011zu,Aguilar-Arevalo:2016ndq}, 
using so-called ``Skipper CCDs'', could allow them to reduce their threshold to near the band gap of silicon~\cite{EstradaTiffenberg}.  
The expected exposure is about 100~gram-years, less than the expected SuperCDMS exposure for silicon.  
Our projections for SuperCDMS can be rescaled easily to get projections for DAMIC.} 
\end{itemize}
\item {\bf Scintillators (NaI, CsI, and GaAs):}
We show projections for hypothetical future experiments using scintillating targets with sensitivity down to one or more photons. 
Such experiments have been argued also to have great potential to sub-GeV DM scattering off electrons~\cite{Derenzo:2016fse}. 
We make projections for three scintillating targets, sodium iodide, cesium iodide, and gallium arsenide, although other possibilities 
exist~\cite{Derenzo:2016fse}.  
We assume a flat $\beta$-decay background of 350~events/keVee/kg/year as for the silicon projection above.  
We do not consider any x-ray lines activated by cosmogenics.  
We ignore the $\sigma_t$ term in Eq.~(\ref{eq:sigmaee}), and use the values listed in Table~\ref{tab:Eg} 
(see also~\cite{Derenzo:2016fse}). 
\end{itemize}

\mysection{Results}

\begin{figure*}[htbp]
	\centering
	\includegraphics[width=0.49\textwidth]{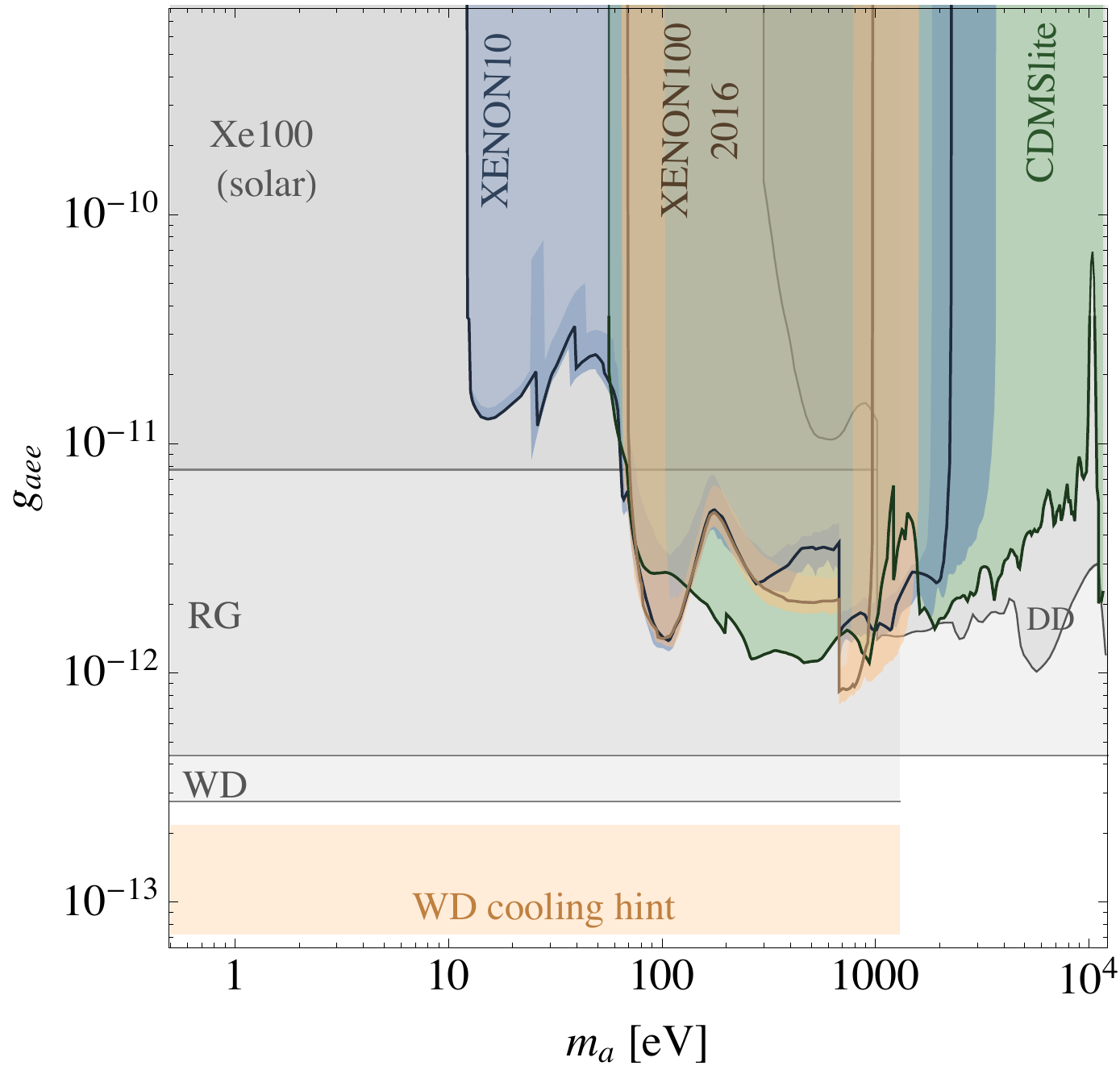}~~~
	\includegraphics[width=0.49\textwidth]{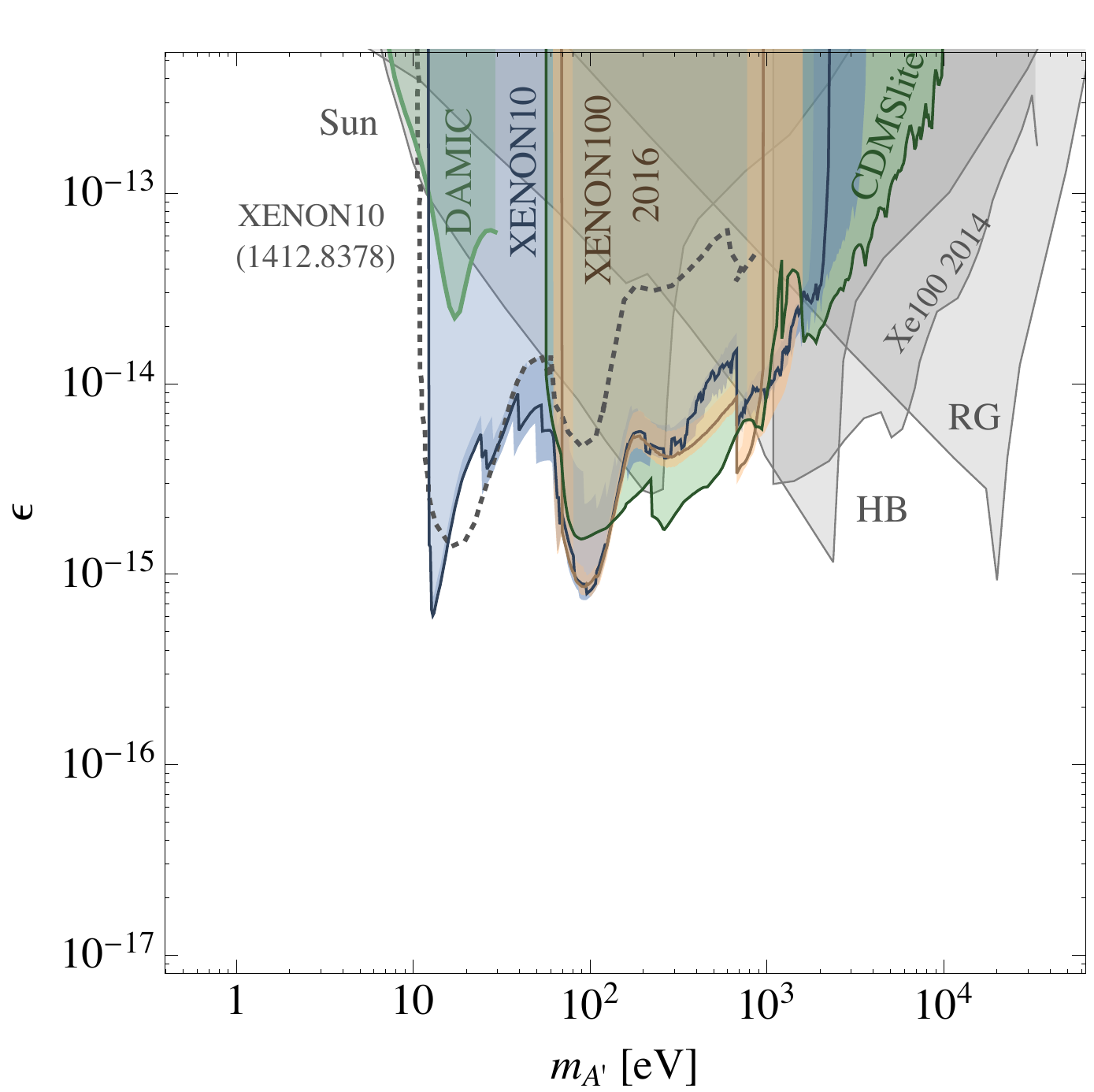}
	\includegraphics[width=0.49\textwidth]{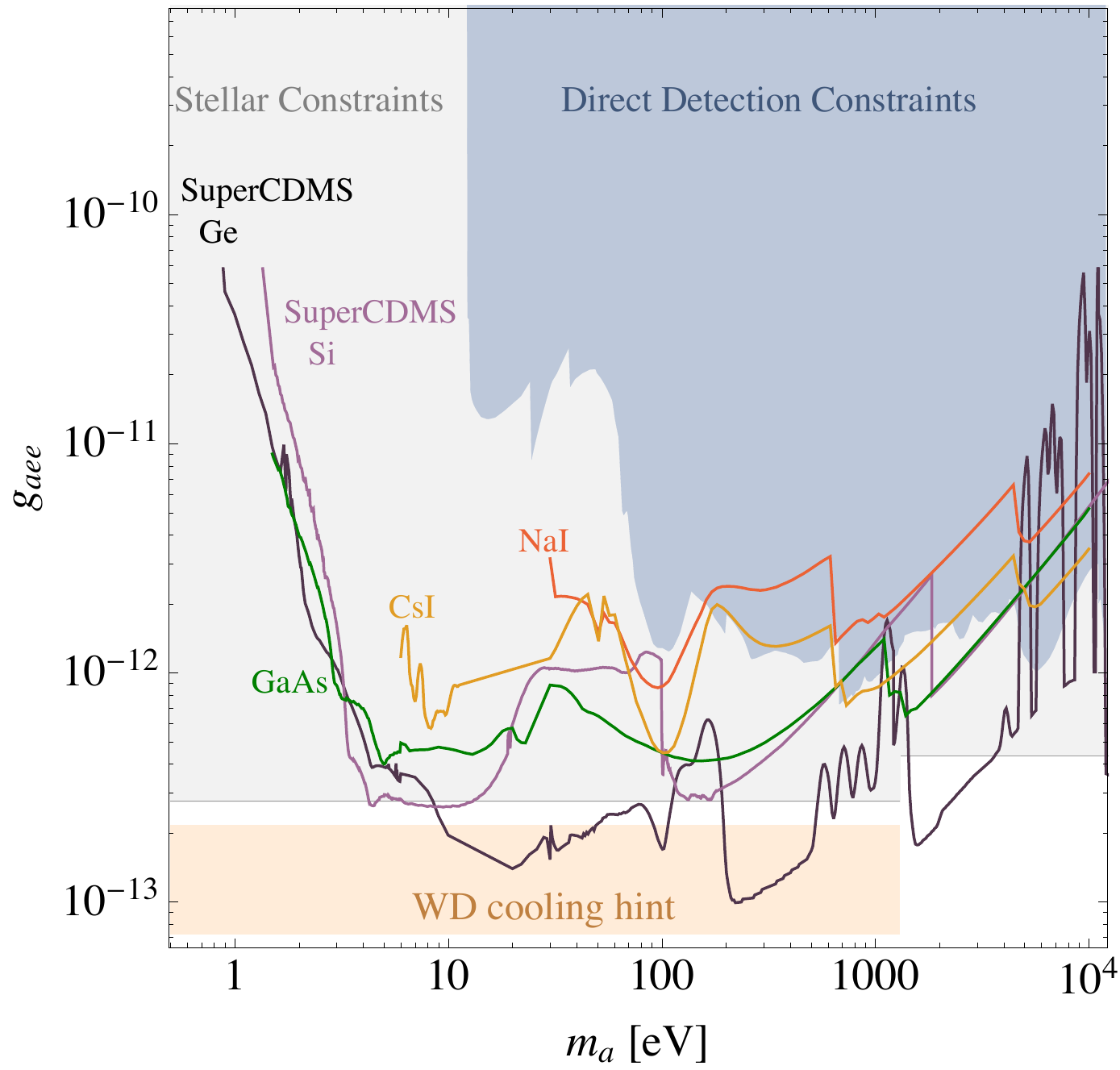}~~~
	\includegraphics[width=0.49\textwidth]{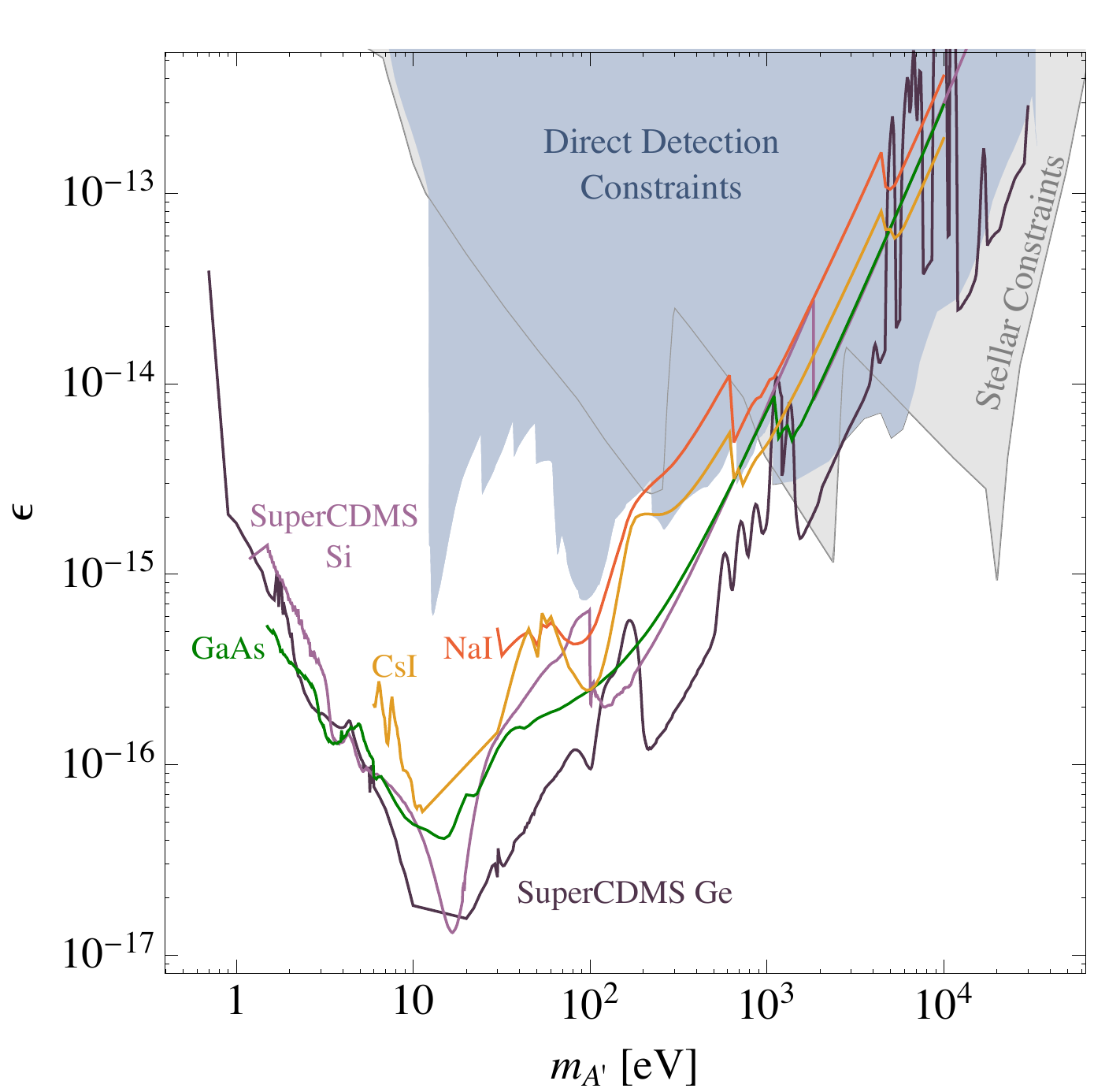}
	\caption{
	Constraints ({\bf shaded regions}) and prospective sensitivities ({\bf solid colored lines}) 
	for axion-like particle (ALP) dark matter ({\bf left}) and dark-photon ($A'$) dark matter ({\bf right}), 
	assuming that the ALP/$A'$ constitutes all the dark matter.  
	{\bf Colored regions} show constraints from XENON10, XENON100, and CDMSlite, as derived in this work, as well as the DAMIC results for $A'$ from ~\cite{Aguilar-Arevalo:2016zop}.
	{\bf Shaded bands} around XENON10 and XENON100 limits show how the bound varies when changing the modeling of the 
	secondary ionization in xenon. 
	{\bf Deep- and light-purple solid lines} show projected 90\% C.L.~sensitivities for SuperCDMS SNOLAB HV using either 
	Ge (20 kg-years) or Si (10 kg-years) targets, respectively.  
	{\bf Yellow}, {\bf orange}, and {\bf green solid lines} show projected sensitivities for hypothetical experiments with the 
	scintillating targets CsI, NaI, and GaAs, assuming an exposure of 10 kg-years.  
	All projections assume a realistic background model discussed in the text, but zero dark counts to achieve sensitivity to 
	low-energy electron recoils.  
	In-medium effects are included for all $A'$ constraints and projections.  
	Shaded gray regions show known constraints from anomalous cooling of the Sun, red giant stars (RG), white dwarf stars (WD), 
	 and/or horizontal branch stars (HB), which are independent of the ALP or $A'$ relic density.  
	 Also shown ({\bf left}) are the combined bounds from XENON100~\cite{Aprile:2014eoa}, EDELWEISS~\cite{Armengaud:2013rta}, 
	 CDMS~\cite{Ahmed:2009ht}, and CoGeNT~\cite{Aalseth:2008rx}; and ({\bf right}) a bound derived in~\cite{An:2014twa} 
	 based on XENON100 data from 2014~\cite{Aprile:2014eoa}. 
	Shaded orange region in left plot is consistent with an ALP possibly explaining the white dwarf luminosity function. 
\label{fig:ResultsDM}
	}
\end{figure*}

Existing constraints and projected sensitivities are shown for ALP ($A'$) 
DM on the left (right) of Fig.~\ref{fig:ResultsDM}.  
For ALPs, we see that the newly derived direct detection bounds from XENON100 and CDMSlite 
partially improve on published bounds from 
CoGeNT, CDMS, XENON100, EDELWEISS, and 
KIMS~\cite{Aalseth:2008rx,Ahmed:2009ht,Armengaud:2013rta,Aprile:2014eoa,Yoon:2016ogs}.  
However, we see that these are weaker than stellar cooling 
bounds~\cite{An:2013yfc,Agashe:2014kda,Raffelt:1985nj,An:2014twa,Redondo:2013lna,Blinnikov:1994eoa,Bertolami:2014wua}.  
Prospective searches, especially SuperCDMS SNOLAB HV with germanium, could improve by a factor of a few beyond the stellar cooling 
constraints.  Intriguingly, this includes probing part of the region consistent with a possible hint for anomalous energy loss in white 
dwarf stars~\cite{Agashe:2014kda,Bertolami:2014wua,Isern:2008nt,Isern:2008fs}. 
If the ALP mass is less than the kinetic energy of electrons, which is $3T/2\sim1.3$~keV, it will contribute to the cooling of the white dwarf \cite{Raffelt:1985nj, Agashe:2014kda}. 

For $A'$ DM, we have derived several constraints that go beyond the constraints from the anomalous energy loss in the Sun, 
horizontal-branch (HB) stars, and red-giant stars~\cite{An:2014twa} (see also~\cite{An:2013yfc,An:2013yua,Redondo:2013lna}).  
First, we have updated the bound derived in~\cite{An:2014twa} based on the low-threshold XENON10 data~\cite{Angle:2011th}. 
We find that this bound  disfavors a gap of several hundred eV in mass between the Sun and HB cooling bounds.  
Existing CDMSlite and XENON100 data disfavor additional parameter space beyond this. 
Prospective searches by SuperCDMS SNOLAB HV using germanium or silicon targets can probe up to more than an order of magnitude 
in $\epsilon$ beyond existing constraints for sub-keV $A'$ masses.  New experiments using scintillator targets with sensitivity to one or 
a few photons can have similar reach.  

\begin{figure}[htbp]
\begin{center}
\includegraphics[width=0.5\textwidth]{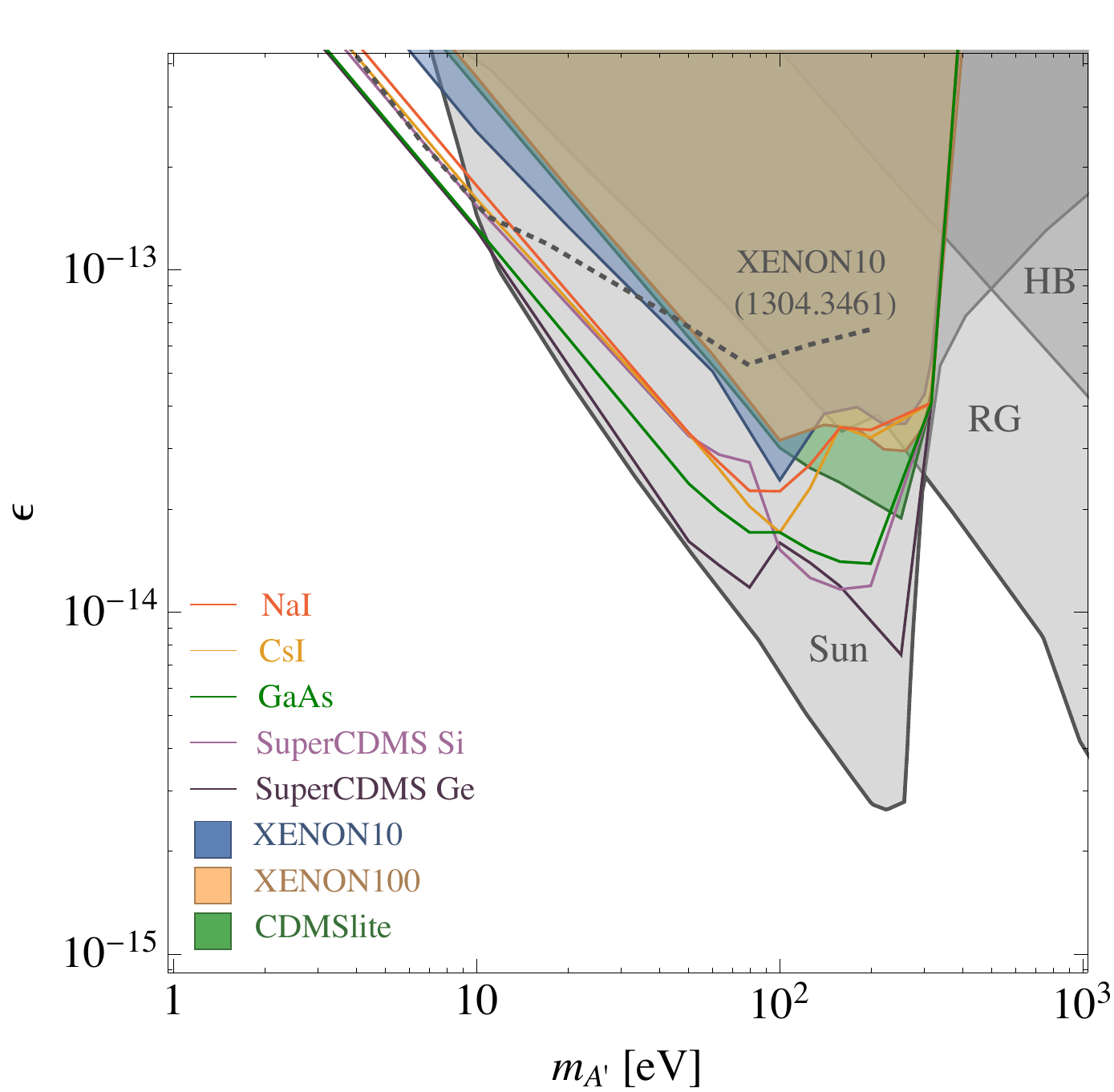}
\caption{
	Constraints ({\bf shaded regions}) and prospective sensitivities ({\bf solid colored lines}) 
	for dark-photons ($A'$) with a St\"uckelberg mass from the Sun via resonant production (including in-medium effects). 
	{\bf Colored regions} show constraints from XENON10, XENON100, and CDMSlite, as derived in this work.  
	{\bf Deep- and light-purple solid lines} show projected 90\% C.L.~sensitivities for SuperCDMS SNOLAB HV using either 
	Ge (20 kg-years) or Si (10 kg-years) targets, respectively.  
	{\bf Yellow}, {\bf orange}, and {\bf green solid lines} show projected sensitivities for hypothetical experiments with the 
	scintillating targets CsI, NaI, and GaAs, assuming an exposure of 10 kg-years.  
	All projections assume a realistic background model discussed in the text, but zero dark counts to achieve sensitivity to 
	low-energy electron recoils.  
	Shaded gray regions show constraint from anomalous cooling of the Sun. 
	 Also shown ({\bf dotted line}) is the bound derived in~\cite{An:2013yua} based on XENON10 data~\cite{Aprile:2014eoa}. 
\label{fig:solar_aprime}
\vspace{-4mm}
}
\end{center}
\end{figure}

We show the results for the solar $A'$ in Fig.~\ref{fig:solar_aprime}.  We find that existing constraints from XENON10, XENON100, 
and CDMSlite are weaker than the stellar cooling constraints~\cite{An:2013yfc, Redondo:2013lna,Agashe:2014kda,Raffelt:1985nj}.
We checked that data from KIMS~\cite{Yoon:2016ogs}, in the absence of a background model, provides a weaker bound (not shown) than CDMSlite. 
However, the future SuperCDMS, as well as possible scintillating target experiments, can significantly improve on existing bounds 
for $A'$ masses below $\sim10$ eV. 

Significant progress is being made in lowering the threshold of DM direct detection experiments.  In this work, we studied 
the implications of this progress for searching for dark absorption.  
We find that planned and possible new experiments, using semiconducting and scintillating targets,  
can significantly improve on existing constrains, even probing well beyond astrophysical bounds.  
This provides unprecedented sensitivity to low-mass dark photons and ALPs. 
\clearpage

\mysection*{Note added:}
While completing this work, we became aware of~\cite{Hochberg:2016sqx}, which considers related topics. 

\mysection*{Acknowledgments}

We thank Haipeng An, Ranny Budnik, Lauren Hsu, Maxim Pospelov, Josef Pradler, and Javier Redondo for useful discussions or correspondence.  
R.E.~is supported by the DoE Early Career research program DESC0008061 and through a Sloan Foundation Research Fellowship. 
T.-T.Y.~is also supported by grant DESC0008061. 
T.V. is supported by the European Research Council (ERC) under the EU Horizon 2020 Programme (ERC-CoG-2015 - Proposal n. 682676 LDMThExp), by the PAZI foundation, by the German-Israeli Foundation (grant No. I-1283- 303.7/2014) and by the I-CORE Program of the Planning Budgeting Committee and the Israel Science Foundation (grant No. 1937/12).
\appendix
\mysection*{APPENDICES}

\begin{figure*}[h]
	\centering
	\includegraphics[width=0.48\textwidth]{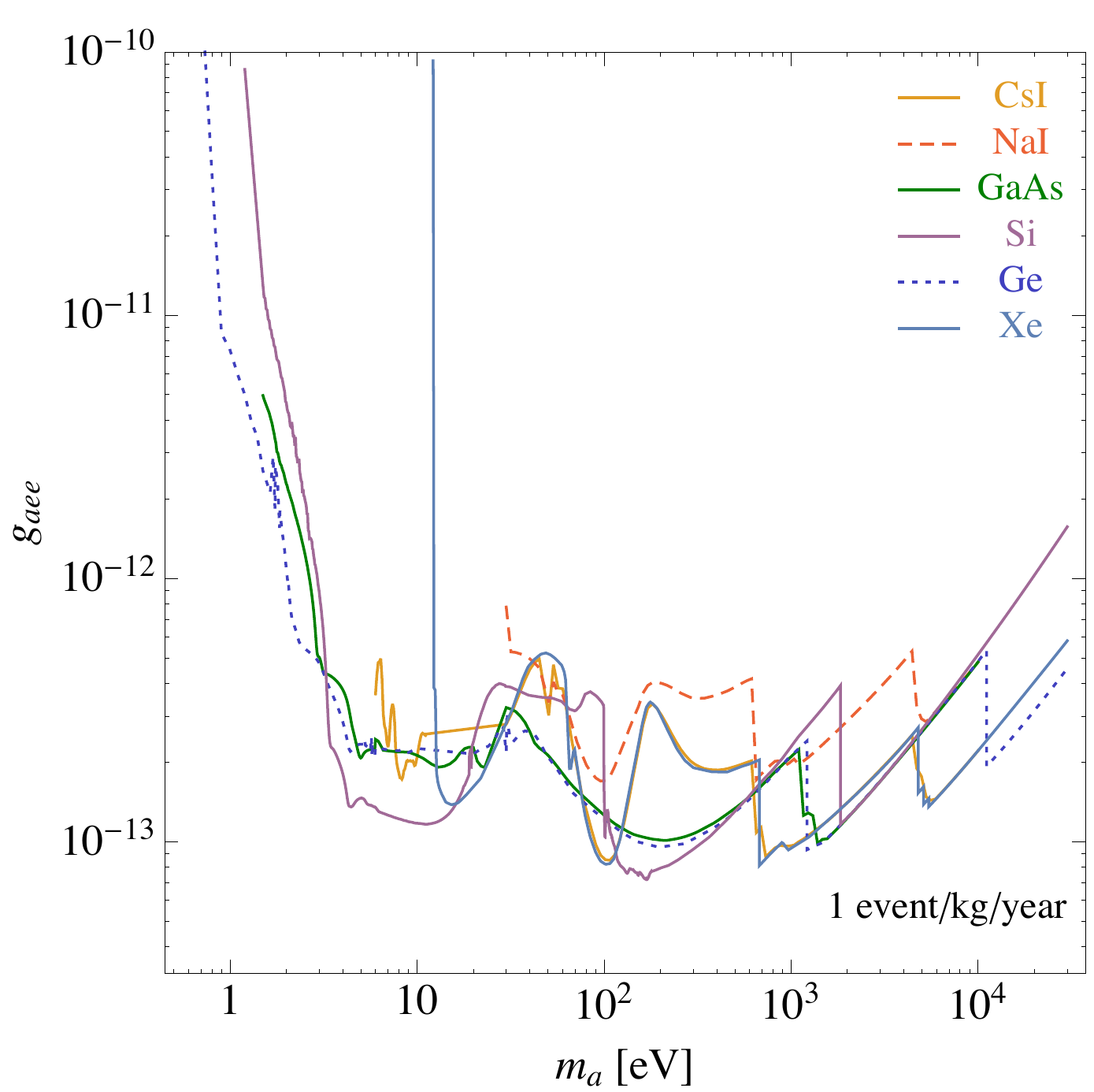}~~
	\includegraphics[width=0.48\textwidth]{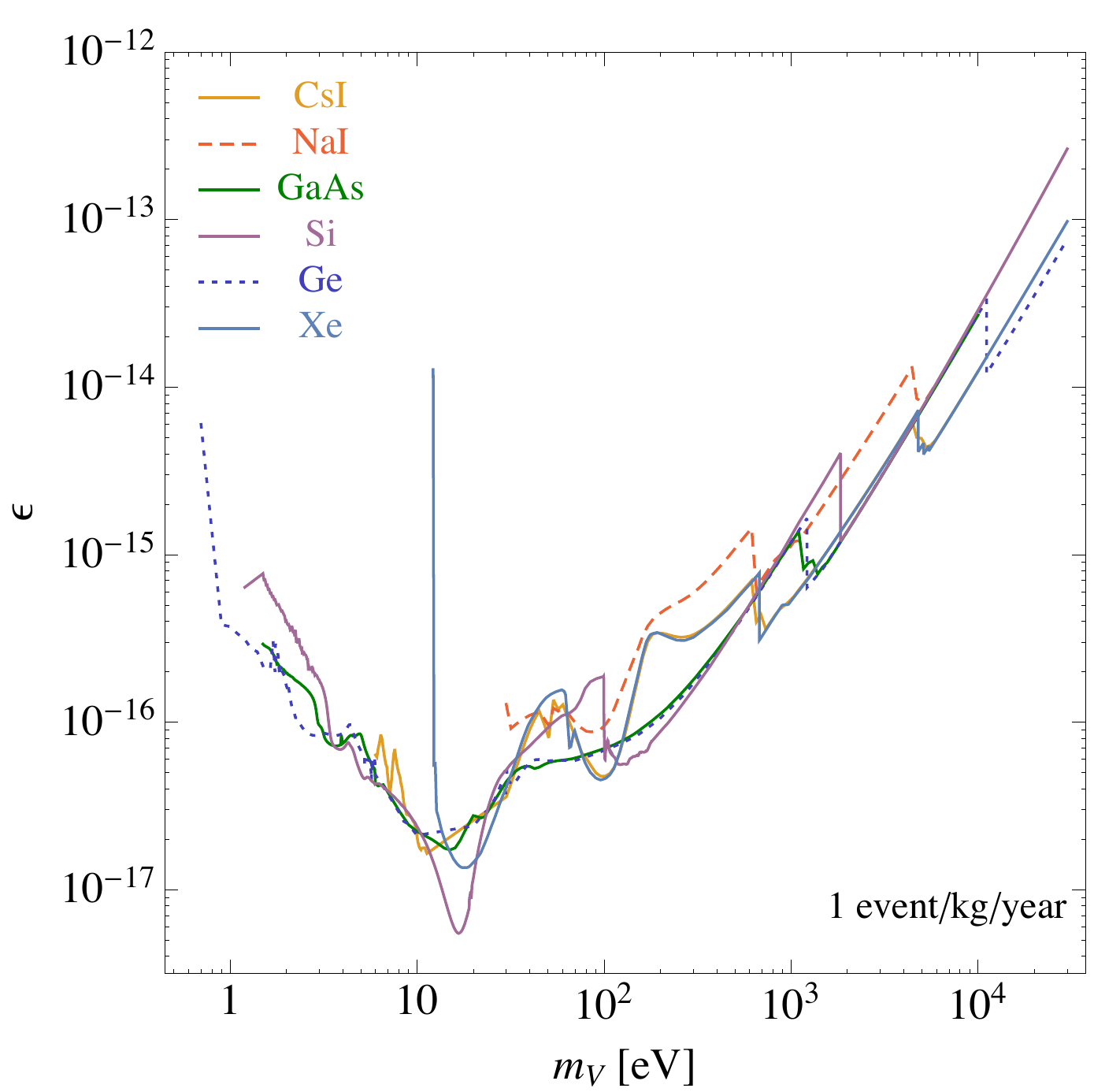}
	\caption{Couplings needed for 1 event/kg/year for ALPS ({\bf left}) and $A'$ ({\bf right}). GaAs and Ge, and Xe and CsI have very similar behaviors due to similar refractive indices. 
	\label{fig:DMrates}
	}
\end{figure*}

\mysection{Rates for ALPs and dark photons in various materials}\label{app:rates}
In Fig.~\ref{fig:DMrates}, we show the couplings $\epsilon$ and $g_{aee}$ for detecting 1 event/kg/year in a variety of materials for the $A'$ and ALP models, respectively, as a function of mass. Note that these calculations do not depend on any background model.  
Therefore, one can easily translate these rates into a projection for a given background model or a limit for a given set of experimental data.

\mysection{Solar $A'$ Production}
A detailed description of the solar dark photon production and absorption rate can be found in Ref.~\cite{An:2013yua}. We describe the relevant equations for the Stueckelberg case here.

The production rate is given by
\beq
\frac{d\Gamma_{T,L}^{prod,V}}{d\omega dV}=
\epsilon_{T,L}^2\frac{d\Gamma_{T,L}^{prod}}{d\omega dV}
\eeq
where $\epsilon_{T,L}$ is the effective mixing angle defined in Eq.~(\ref{eq:effective_epsilon}). 
Inside the Sun the photon self energy is given by 
\bea
{\rm{Re}}~\Pi_T&=& \omega_p^2, \quad
{\rm{Re}}~\Pi_L= \omega_p^2 (m_{A'}^2/\omega^2), \\
{\rm{Im}}~\Pi_{T,L}&=& -\omega(1-e^{-\omega/T})\Gamma_{T,L}^{abs},
\eea
where $\omega_p(r)={e^2 n_e(r)}/{m_e}$ is the plasma frequency inside the Sun, $n_e$ is the electron density inside the Sun, and $m_e$ is the electron mass ~\cite{Redondo:2008aa}. There are two primary production modes for the dark photon: resonant production and bremsstrahlung. 
Below the maximum plasma frequency $\omega_p^{max}\simeq 300\rm~eV$, the dominant contribution to the flux is longitudinal resonant production. 
The resonant production rate is given as 
\begin{align}
\left.\frac{d\Gamma_{T,L}^{prod,V}}{d\omega dV}\right|_{res}
=\frac{\epsilon^2 m_{A'}^2\sqrt{\omega^2-m_{A'}^2}}{2\pi(1-e^{\omega/T})}
\delta(m_{A'}^2- \rm{Re}~\Pi_{T,L})
\end{align}
%
Using Eq.~(\ref{RatetoFlux}), we obtain the corresponding flux from the resonant production,
\bea
\left.\frac{d\Phi_{L}}{d\omega}\right|_{res}&=&\frac{1}{4\pi d_\odot^2} \frac{\epsilon^2m_{A'}^2 r^2(\omega_p) \omega^3}{e^{\omega/T(\omega_p)}-1} \left.\frac{2}{d\omega_p^2/dr}\right|_{\omega_p=\omega} 
\\
\left.\frac{d\Phi_{T}}{d\omega}\right|_{res}&=&\frac{\epsilon^2}{4\pi d_\odot^2}  \frac{ m_{A'}^4 r^2(\omega_p) \sqrt{\omega^2-m_{A'}^2}}{e^{\omega/T(\omega_p)}-1} \left.\frac{2}{d\omega_p^2/dr}\right|_{\omega_p=m_{A'}}.
\eea

For $\omega>\omega_p^{max}$ since the resonant production is absent , bremsstrahlung is the dominant production mode, 
\begin{align}
&\left.\frac{d\Gamma_{L}^{prod,V}}{d\omega dV}\right|_{brem} =\sum_{i=e,\rm H,He} \epsilon_{T,L}^2\frac{8Z_i^2 \alpha^3 n_e n_i m_V^2 \sqrt{\omega^2-m_{A'}^2}}{3m_e^2\omega^4}
\sqrt{\frac{8m_e}{\pi T}} f\left(\sqrt{\frac{\omega}{T}}\right)
\end{align}
where 
\begin{align}
f(a)=\int_a^\infty dx\ x e^{-x^2} {\rm log }\left(\frac{x+\sqrt{x^2-a^2}}{x+\sqrt{x^2-a^2}} \right).
\end{align}
For H and He, we take the solar data from~\cite{PenaGaray:2008qe}. 
Note that there is a suppression of $\alpha^3$ and thermal distribution ($e^{-\omega/T}$) to the bremsstrahlung production mode for $\omega>300~\rm eV$.

\bibliography{LDMmodels}
\bibliographystyle{apsrev}

\end{document}